\newcommand{\be}{\begin{equation}}
\newcommand{\ee}{\end{equation}}
\newcommand{\bea}{\begin{eqnarray}}
\newcommand{\eea}{\end{eqnarray}}
\newcommand{\nn}{\nonumber\\ }
\newcommand{\p}[1]{(\ref{#1})}

\newcommand{\cD}{{\cal D}}
\newcommand{\cDb}{{\bar{\cal D}}}
\newcommand{\Db}{{\bar D}}
\newcommand{\tb}{{\bar\theta}}
\newcommand{\xib}{{\bar\xi}}

\newcommand{\bF}{{\bar F}}
\newcommand{\Dt}{{\mbox{\tt D}}}
\newcommand{\vfi}{{\varphi}}
\newcommand{\bvfi}{{\bar\varphi}}
\newcommand{\ve}{{\varepsilon}}
\documentclass[12pt]{article}
\usepackage{amscd,amsmath,amssymb}

\begin{document}
\renewcommand{\thefootnote}{\fnsymbol{footnote}}
\begin{titlepage}
\begin{flushright}
hep-th/0212295\\
%\today
\end{flushright}
\vskip 0.6truecm
\begin{center}
{\Large\bf
 Goldstone Superfield Actions in AdS$_{\bf 5}$ backgrounds}
\end{center}
 \vskip 0.6truecm
 \begin{center}
{\large\bf S. Bellucci}\footnote{bellucci@lnf.infn.it } \vspace{0.5cm} \\
{\it INFN-Laboratori Nazionali di Frascati, \\
 C.P. 13, 00044 Frascati, Italy \vspace{0.5cm} }\\
{\large\bf E. Ivanov\footnote{eivanov@thsun1.jinr.ru }
and S. Krivonos}\footnote{krivonos@thsun1.jinr.ru }
  \vspace{0.5cm} \\
{\it Bogoliubov  Laboratory of Theoretical Physics, JINR,\\
141980 Dubna, Moscow Region, Russia} \vspace{1.5cm}
\end{center}
\vskip 0.6truecm  \nopagebreak
\begin{abstract}
\noindent Nonlinear realizations superfield techniques,
pertinent to the description of partial breaking of global $N=2$
supersymmetry in a flat $d=4$ super Minkowski background, are generalized to the case of partially
broken $N=1$ AdS$_5$ supersymmetry $SU(2,2|1)$.
We present, in an explicit form, off-shell manifestly
$N=1, d=4$  supersymmetric
minimal Goldstone superfield actions for two patterns of partial
breaking of $SU(2,2|1)$ supersymmetry. They correspond
to two different nonlinear realizations of the latter, in the supercosets with
the AdS$_5$ and AdS$_5\times S^1$ bosonic parts. The relevant worldvolume
Goldstone supermultiplets are accommodated, respectively, by improved tensor and
chiral $N=1, d=4$ superfields. The second action is obtained from the first one by
dualizing the improved tensor Goldstone multiplet into a chiral Goldstone one.
In the bosonic sectors, the first and second actions yield static-gauge Nambu-Goto
actions for a L3-brane on AdS$_5$ and a scalar 3-brane on AdS$_5\times S^1$.

\end{abstract}
\newpage

\end{titlepage}

\renewcommand{\thefootnote}{\arabic{footnote}}
\setcounter{footnote}0
\setcounter{equation}0
\section{Introduction}
The description of superbranes in terms of worldvolume Goldstone superfields
based on the concept of partial spontaneous breaking of global supersymmetry
(PBGS) \cite{bw,pbgs} is advantageous in many respects. Its main attractive
feature is that the corresponding invariant actions reveal manifest
off-shell linearly realized worldvolume supersymmetry. The second half of
the full supersymmetry is nonlinearly realized \cite{BG1}-\cite{bik1B} (see \cite{bik0,i1}
for a review and further references). A group-theoretical basis of the PBGS theories
is provided by the appropriate versions of the general nonlinear
realizations method \cite{nonl}. The PBGS approach has many potential capabilities and
implications in string theory, e.g. for constructing non-abelian Born-Infeld actions,
as well as their supersymmetric extensions, for describing different possibilities of a
non-standard partial supersymmetry breaking (such as $1/4, 3/4, \ldots $)
and studying the Hamiltonian and quantum structure of the relevant
models, etc (see e.g. \cite{bgik1}-\cite{ikn}).
%In particular, the quantum mechanics of a massive superparticle
%in $d=4$, which preserves 1/4 of the target space supersymmetry with eight
%supercharges, has been studied in \cite{bgik1}. The latter theory
%corresponds to the partial breaking of $N=8, d=1$ supersymmetry  down to
%$N=2$. Also, the Hamiltonian of a $N=4$ superparticle on $AdS_2\times S^2$
%background has been recently constructed \cite{bgik2}, taking advantage of
%the classical equivalence between a simple (super)conformal mechanics model and a
%charged massive (super)particle propagating near the $AdS_2\times S^2$ horizon of
%an extreme Reissner-Nordstr\"om black hole. This equivalence \cite{bik12,ikn} was established
%within the appropriate versions of the general nonlinear realizations approach \cite{nonl}
%which constitutes a basis of the PBGS theories.

Most PBGS theories constructed so far correspond to superbranes on flat
super Minkowski backgrounds.
On the other hand, in the framework of the AdS/CFT correspondence \cite{ads},
of primary interest are the AdS$\times S$ and pp-wave
type \cite{mets} superbackgrounds, with $S$ being some curved
Riemannian manifold, e.g. a sphere $S^n$. While Green-Schwarz type actions
for branes on such backgrounds are known (see e.g. \cite{mt1,mt2}), not too many explicit
examples of the worldvolume superfield PBGS actions were presented.
Until now such actions were given only for $N=1$ supermembrane in AdS$_4$ \cite{dik},
$N=1$ superstring in AdS$_3$ \cite{ik} and $N=2$ superparticle in AdS$_2$ \cite{ikn}
(the two latter examples are dimensional reductions of the first one). The corresponding
groups of superisometries coincide with superconformal groups in dimensions lower by 1
(in $d=3$, $d=2$ and $d=1$), so the construction of these PBGS systems
amounts to setting up appropriate nonlinear realizations of superconformal symmetries.

It is tempting to construct PBGS versions of superstring and D3-brane on
the AdS$_5\times S^5$ superbackground which is the basic ingredient
of the original AdS/CFT conjecture. They should reappear in this
context as theories of partial breaking of the $N=4$ superconformal
group $SU(2,2|4)$. It is natural to begin with some their truncations,
corresponding to nonlinear realizations of the simpler cases of $N=1$ and $N=2$, $d=4$
superconformal groups $SU(2,2|1)$ and $SU(2,2|2)$. In \cite{kum}
an attempt to construct a PBGS model for $SU(2,2|1)$ was undertaken.
This model generalizes that of \cite{BG2} and should reduce to it in the limit of infinite
AdS$_5$ radius $R$. Similarly to the model of \cite{BG2}, also its generalization
employs a $N=1$ chiral superfield as the basic
Goldstone one and is expected to describe a scalar 3-brane on AdS$_5\times S^1$.
However, the proper minimal Goldstone superfield action
was not constructed in \cite{kum}, though it was suggested that such
an action could be regained like this has been done in the flat case in \cite{BG2,RT,R2}.
There, an alternative PBGS action with a linear (tensor)
$N=1$ Goldstone multiplet was firstly constructed and then it was dualized
into an action of $N=1$ chiral Goldstone supermultiplet.

The basic aim of the present paper is to carry out such
a construction in the case of AdS$_5$
background.\footnote{Its sketchy exposition was given in \cite{bik02}.}
Instead of dealing with a nonlinear realization of $SU(2,2|1)$
in the standard approach \cite{nonl}, we follow the line of
refs. \cite{BG2,RT,R2,dik,ik1}. We firstly define a non-standard $N=2, d=4$ superspace
of $SU(2,2|1)$ which contains  Grassmann coordinates associated with both
Poincar\'e and conformal supersymmetry generators. Then we seek for
a representation of $SU(2,2|1)$ on the chiral and improved tensor
$N=1$ superfields defined as $N=1$ components of a suitable
`$N=2, d=4$ tensor multiplet' with a Goldstone-type transformation law.
We find that requiring the closure of $SU(2,2|1)$ on this set of $N=1$
superfields necessarily implies the constraints, which are a generalization of
those used in \cite{BG2,RT,R2}. They allow one to trade the
chiral superfield for the improved tensor one and to construct
a $SU(2,2|1)$ invariant action of the tensor $N=1$ Goldstone superfield.
The action describes a AdS$_5$ L3-superbrane \footnote{See \cite{howe} for the relevant
nomenclature.} in a static gauge and goes into
that of refs. \cite{BG2,RT,R2}, upon taking the $R=\infty$ limit.
Then we dualize the Goldstone tensor multiplet into a chiral one and
thus obtain the Goldstone superfield action with the AdS$_5\times S^1$ 3-brane
bosonic part, i.e. just that corresponding to the PBGS option analyzed
in ref. \cite{kum}.

This paper is organized as follows. In Section 2 we recall the basic
properties and existing superfield formulations of the $N=2, d=4$ tensor multiplet
which plays a central role in our construction. In Section 3
we repeat, in a somewhat different setting, the construction of the PBGS
superfield action for the $N=1$ Goldstone tensor multiplet in
a flat $d=4$ super Minkowski background.
We demonstrate that the requirement of
covariance under the nonlinearly realized $d=5$ Lorentz $SO(1,4)$ symmetry
provides an alternative way of deducing the constraints which should be imposed \cite{BG2}
on the linear tensor $N=2$ supermultiplet in order to obtain the appropriate
nonlinear realization in terms of the single Goldstone tensor
$N=1$ supermultiplet.
In Section 4 we generalize the whole construction to the
AdS$_5$ case, corresponding to the one-half partial breaking of $SU(2,2|1)$
supersymmetry (down to $N=1, d=4$ Poincar\'e supersymmetry).
For self-consistency, in this case we are led to impose from the very beginning the constraints which
properly generalize those of the flat background. The resulting
nonlinear realization is the genuine generalization of the flat case
one, and the $SU(2,2|1)$ invariant action of the Goldstone improved
tensor $N=1$ superfield can be constructed in a similar fashion.
In Section 5 we perform a duality transformation of the improved tensor Goldstone
$N=1$ superfield action we have constructed and arrive at the invariant action of a chiral
Goldstone $N=1$ superfield. The latter properly generalizes the minimal Goldstone superfield
action of refs. \cite{BG1,BG2}. An outline of some open problems is the
content of the concluding Section.

\setcounter{equation}0
\section{N=2 tensor multiplet}
Since the central role in our study is played by a proper
generalization of the $N=2$ tensor multiplet, we start with a brief
recapitulation of the basic properties of the multiplet and its
different formulations.

In the standard $N=2, d=4$ superspace $z =(x^{\alpha\dot\alpha}, \theta^\alpha_i,
\bar\theta^{\dot\alpha\,i})$ the $N=2$ tensor multiplet is described by an
isotriplet of scalar fields $L^{ij}(z)$ satisfying the constraints \cite{tm}
\be\label{tmdef1}
\cD^{(i}_{\alpha} L^{jk)}=0\;, \quad \cDb_{\dot\alpha}^{(i} L^{jk)}=0 \;, \quad
\left( L^{ij} \right)^\dagger =-L_{ij}\equiv -\epsilon_{ik}\epsilon_{jl}L^{kl} \;,
\ee
where\footnote{We mostly use the conventions of \cite{book}.}
\be\label{n2cd}
\cD^i_\alpha=\frac{\partial}{\partial \theta^\alpha_i}+
i\tb^{\dot\alpha i}\partial_{\alpha\dot\alpha} \;,\quad
\cDb_{\dot\alpha i}=-\frac{\partial}{\partial\tb^{\dot\alpha i}}-
i\theta^\alpha_i\partial_{\dot\alpha \alpha} \;, \quad
\left( \theta^{\alpha}_i \right)^\dagger =\tb^{\dot\alpha\,i}\;.
\ee
It is convenient to introduce two sets of Grassmann coordinates and covariant derivatives
\bea\label{n1cd}
&& \left( \theta_{\alpha 1}, \tb_{\dot\alpha}^1\right)
\equiv \left( \theta_\alpha ,\tb_{\dot\alpha}\right), \;
\left( \cD^1_\alpha , \cDb_{\dot\alpha 1} \right) \equiv \left( D_\alpha ,\Db_{\dot\alpha} \right), \nn
&&\left( \theta_{\alpha 2} ,\tb_{\dot\alpha}^2\right) \equiv \left( \xi_\alpha ,{\bar\xi}_{\dot\alpha} \right) \;, \quad
\left( \cD^2_\alpha , \cDb_{\dot\alpha 2} \right) \equiv \left( \nabla_\alpha ,{\bar\nabla}_{\dot\alpha} \right) \;,
\eea
and rewrite \p{tmdef1} as follows:
\be\label{tmdef2}
\left\{ \begin{array}{l}
D_\alpha F =0 \;, \\
D_\alpha L -\nabla_\alpha  F =0 \;, \\
\nabla_\alpha L + D_\alpha {\bar F} =0 \;, \\
\nabla_\alpha {\bar F}=0 \;,
\end{array}  \right. \qquad
\left\{ \begin{array}{l}
{\bar\nabla}_{\dot\alpha}  F =0 \;, \\
{\bar\nabla}_{\dot\alpha} L +\Db_{\dot\alpha}  F =0 \;, \\
\Db_{\dot\alpha} L - {\bar\nabla}_{\dot\alpha} {\bar F} =0 \;, \\
\Db_{\dot\alpha} {\bar F}=0 ,
\end{array}  \right.
\ee
where
\be
L \equiv 2L^{12} = \bar L\;,\quad  F \equiv L^{11} \;,\quad {\bar F}\equiv -L^{22} \;.
\ee
One observes that the covariant derivatives of all superfields
$\left\{ L, F,{\bar F}\right\}$ with respect to $\xi,\bar\xi$ are expressed from \p{tmdef2}
as covariant derivatives with respect to $\theta,\tb$. Therefore, the only independent
$N=1$ superfield components of these $N=2$ superfields are the
$\xi = \bar\xi=0 $ components of the latter subjected to the constraints (which
also follow from \p{tmdef2})
\be\label{flatcon11}
\Db_{\dot\alpha} {\bar F}=0 \; ,\; D_\alpha  F =0 \;, \; D^2L=\Db^2 L=0~.
\ee
Hence, they are the $N=1$ chiral $(F,{\bar F})$ and tensor $( L)$ superfields.

Two other ways to describe the $N=2$ tensor multiplet are the formulations in
$N=2$ harmonic \cite{hss,book} and projective \cite{pss} superspaces.
Such formulations are basically equivalent, as shown in \cite{kuzenko}. However, the
projective superspace formulation is more suitable for our eventual purpose
of constructing PBGS $N=1$ Goldstone superfield actions, because it allows one to pass easily
to $N=1$ superfield formulations.

Projective superspace includes one additional complex bosonic
$\mathbb{C}\mbox{P}^1$ coordinate $\omega$. One defines
\bea\label{pcd}
&& \cD^{\omega}_\alpha = \omega D_\alpha -\nabla_\alpha\;,\quad
 \cDb^{\omega}_{\dot\alpha}=\Db_{\dot\alpha}+\omega {\bar\nabla}_{\dot\alpha}\;, \\
&& \left\{ \cD^{\omega}_\alpha, \cD^{\omega}_\beta \right\} =0 \;, \;
\left\{ \cD^{\omega}_\alpha, \cDb^{\omega}_{\dot\alpha} \right\}=0 \;, \;
\left\{ \cDb^{\omega}_{\dot\alpha}, \cDb^{\omega}_{\dot\beta} \right\} =0 \;,
\eea
and rewrites the original constraints \p{tmdef1} as
\be\label{htm5}
\cD^{\omega}_\alpha {\cal L}^\omega =0 \; \quad \cDb^{\omega}_{\dot\alpha} {\cal L}^\omega =0 \;,
\ee
where
\be\label{l}
{\cal L}^\omega =\omega F +L -\frac{\bar F}{\omega} \;.
\ee
Thus, the components of ${\cal L}^\omega$ are effectively $N=1$ superfields and one can
construct $N=2$ invariants which look much like $N=1$ ones and contain
integration only over $N=1$ superspace.

\setcounter{equation}0
\section{Goldstone  N=1 tensor multiplet in a flat background}
The idea to utilize the $N=1$ tensor multiplet as the Goldstone one for describing
the partial breaking of global $N=2, d=4$ Poincar\'e supersymmetry down to
$N=1$ has been worked out in \cite{BG2,RT,R2}. The basic ingredients
of this construction are, first of all, an appropriate modification of the linear
$N=2$ tensor multiplet and, secondly, the covariant constraints imposed
on the latter, in order to end up with a nonlinear realization of the
$N=2 \rightarrow N=1$ PBGS pattern in terms of a single Goldstone
$N=1$ tensor multiplet. The corresponding Goldstone
superfield action is a manifestly $N=1$ supersymmetric worldvolume form of
the action of $N=1, d=5$ L$3$-brane in a flat background. Here we reproduce
this construction in a different setting, which admits a direct
generalization to the case we are mainly interested in, i.e. the
partial breaking of the $SU(2,2|1)$ supersymmetry down to the $N=1, d=4$
Poincar\'e supersymmetry. In passing, we show that
the constraints of refs. \cite{BG2,RT,R2} naturally arise from the
requirement of 5-dimensional Lorentz $SO(1,4)$ covariance.

We basically follow the line of refs. \cite{BG1,BG2}. As a first step, we
should define a `linear' version of $N=2 \rightarrow N=1$ PBGS in $N=2, d=4$
superspace with $N=2$ tensor multiplet as the relevant Goldstone one.
We start with $N=2, d=4$ Poincar\'{e} superalgebra extended by
a real central charge $D$
\be\label{flatalgebra}
 \left\{ Q_{\alpha},{\bar Q}_{\dot\alpha} \right\}=2P_{\alpha\dot\alpha}\;,\;
 \left\{ S_{\alpha},{\bar S}_{\dot\alpha} \right\}=2P_{\alpha\dot\alpha}\;,\;
\left\{ Q_{\alpha},S_{\beta} \right\}=-\ve_{\alpha\beta}D \;,\;
 \left\{ {\bar Q}_{\dot\alpha},{\bar S}_{\dot\beta} \right\}=-\ve_{\dot\alpha\dot\beta}D \;.
\ee
Here $Q_{\alpha},{\bar Q}_{\dot\alpha}$ and $S_{\alpha},{\bar S}_{\dot\alpha}$ are generators
of unbroken and broken $N=1$ supersymmetries, respectively. The latter generators
and the 4-translation generator $P_{\alpha\dot\alpha}$ possess standard
commutation relations with the Lorentz $so(1,3)$ generators
$( M_{\alpha\beta},{\bar M}_{\dot\alpha\dot\beta})$
\bea
&& i\left[  M_{\alpha\beta},M_{\rho\sigma}\right] =
 \ve_{\alpha\rho}M_{\beta\sigma}+\ve_{\alpha\sigma}M_{\beta\rho}+
         \ve_{\beta\rho}M_{\alpha\sigma}+
           \ve_{\beta\sigma}M_{\alpha\rho} \equiv
       \left( M \right)_{\alpha\beta,\rho\sigma}\;,\nn
&&i\left[  {\bar M}_{\dot\alpha\dot\beta},{\bar M}_{\dot\rho\dot\sigma}\right]
 = \left( {\bar M} \right)_{\dot\alpha\dot\beta,\dot\rho\dot\sigma}\;,\quad
 i\left[  M_{\alpha\beta},P_{\rho\dot\rho}\right] =
\ve_{\alpha\rho}P_{\beta\dot\rho}+\ve_{\beta\rho}P_{\alpha\dot\rho}\;,\nn
&& i\left[  {\bar M}_{\dot\alpha\dot\beta},P_{\rho\dot\rho}\right] =
\ve_{\dot\alpha\dot\rho}P_{\rho\dot\beta}+\ve_{\dot\beta\dot\rho}P_{\rho\dot\alpha}\;,\quad
 i\left[ M_{\alpha\beta}, Q_{\gamma}\right]=\ve_{\alpha\gamma}Q_{\beta}+
  \ve_{\beta\gamma}Q_{\alpha} \equiv \left( Q\right)_{\alpha\beta,\gamma}\;,\nn
&& i\left[ M_{\alpha\beta}, S_{\gamma}\right]= \left( S\right)_{\alpha\beta,\gamma}\;,\quad
  i\left[ {\bar M}_{\dot\alpha\dot\beta}, {\bar Q}_{\dot\gamma}\right]=
  \left( {\bar Q}\right)_{\dot\alpha\dot\beta,\dot\gamma}\;,\quad
 i\left[ {\bar M}_{\dot\alpha\dot\beta}, {\bar S}_{\dot\gamma}\right]=
  \left( {\bar S}\right)_{\dot\alpha\dot\beta,\dot\gamma}\;.
\eea

Our basic superfield is ${\cal L}^\omega$, eq. \p{l}.
We associate it with the generator $D$ as the relevant coset parameter, i.e.
as a Goldstone $N=2$ superfield.
Together with the $N=2$ superspace coordinates
$\left\{ x^{\alpha\dot\alpha}, \theta^\alpha, \tb_{\dot\alpha},\xi^\alpha,\xib_{\dot\alpha}\right\}$,
the latter parameterizes the coset space of $N=2$  Poincar\'{e} supergroup over
its $d=4$ Lorentz subgroup $SO(1,3)$
\be\label{g}
g=e^{-ix^{\alpha\dot\alpha}P_{\alpha\dot\alpha}+i\theta^{\alpha}Q_{\alpha}+
 i{\bar\theta}_{\dot\alpha}{\bar Q}^{\dot\alpha}} e^{i\xi^{\alpha}S_{\alpha}+
 i{\bar\xi}_{\dot\alpha}{\bar S}^{\dot\alpha}}e^{i{\cal L}^\omega D} \; .
\ee
The $\mathbb{C}\mbox{P}^1$ co-ordinate $\omega$ on which ${\cal L}^\omega$ depends can be
regarded, like harmonic coordinates, to parameterize a coset of the $R$-symmetry group $SU(2)$.
However, this internal symmetry is explicitly broken in
the $N=2 \rightarrow N=1$ PBGS actions \cite{BG2}. For this reason in what follows
we shall not be interested in it. We treat $\omega$ as an external parameter,
which allows one to incorporate all  superfield components of the $N=2$ tensor
multiplet into the coset geometry. Thus, in the present case, we are dealing with a special
coset realization of $N=2$ supersymmetry. We call it `linear'
because, though  the $S$-supersymmetry and $D$-transformations of the Goldstone
superfield ${\cal L}^\omega$ are inhomogeneous, they do not exhibit any nonlinearity.

The full set of $N=2$  super Poincar\'{e} transformations of the coset parameters in \p{g}
can be found by acting on \p{g} from the left by various group elements. Most relevant
are the $Q$ and $S$ supersymmetry transformations.

Unbroken $Q$ supersymmetry
($g_0=  e^{i\epsilon^{\alpha}Q_{\alpha}+i{\bar\epsilon}_{\dot\alpha} {\bar Q}^{\dot\alpha}}$) reads:
\be\label{qsusy}
\delta x^{\alpha\dot\alpha}=-
  i\left( \epsilon^\alpha{\bar\theta}^{\dot\alpha}+
  {\bar\epsilon}^{\dot\alpha}\theta^{\alpha}\right),\;
\delta \theta^{\alpha}=\epsilon^{\alpha}\;,\;
\delta {\bar\theta}_{\dot\alpha}={\bar\epsilon}_{\dot\alpha}\;.
\ee

Broken $S$ supersymmetry ($g_0=
  e^{i\eta^{\alpha}S_{\alpha}+i{\bar\eta}_{\dot\alpha} {\bar S}^{\dot\alpha}}$) reads:
\be\label{ssusy}
 \delta x^{\alpha\dot\alpha} = -i \left( \eta^\alpha \xib^{\dot\alpha}+{\bar\eta}^{\dot\alpha}\xi^\alpha   \right)\;,\quad
 \delta\xi^{\alpha}=\eta^{\alpha}\;,\;
   \delta{\bar\xi}^{\dot\alpha}={\bar\eta}^{\dot\alpha}\;,\quad
\delta {\cal L}^\omega= -i \left( \theta\cdot\eta -{\bar\theta}\cdot{\bar\eta}\right).
\ee
 From \p{qsusy} and \p{ssusy} one observes that the `active' form of the transformations
of our basic superfield ${\cal L}^\omega$ is standard with respect to the $Q$ supersymmetry
\be\label{qtr}
\delta^*_Q {\cal L}^\omega = \epsilon^\alpha Q_\alpha {\cal L}^\omega +
 {\bar\epsilon}_{\dot\alpha}{\bar Q}^{\dot\alpha}{\cal L}^\omega
\ee
and is modified by a $\theta$-dependent shift under $S$ supersymmetry
\be\label{str}
\delta^*_S {\cal L}^\omega = -i \left( \theta\cdot\eta -{\bar\theta}\cdot{\bar\eta}\right) +
  \eta^\alpha S_\alpha {\cal L}^\omega +
   {\bar\eta}_{\dot\alpha}{\bar S}^{\dot\alpha}{\cal L}^\omega \;.
\ee
Here
\be
Q_\alpha=-{\partial\over{\partial\theta^{\alpha}}}+i{\bar\theta}^{\dot\alpha}
 \partial_{\alpha\dot\alpha}\;, \;
{\bar Q}_{\dot\alpha}={\partial\over{\partial{\bar\theta}^{\dot\alpha}}}-
   i{\theta}^{\alpha}\partial_{\alpha\dot\alpha} \;,\;
S_\alpha=-{\partial\over{\partial\xi^{\alpha}}}+i{\bar\xi}^{\dot\alpha}
 \partial_{\alpha\dot\alpha}\;, \;
{\bar Q}_{\dot\alpha}={\partial\over{\partial{\bar\xi}^{\dot\alpha}}}-
   i{\xi}^{\alpha}\partial_{\alpha\dot\alpha} \;.
\ee
This modification, being independent of $\omega$, affects only the transformation law of
$L$, leaving that of $F,{\bar F}$ in its standard form. Nevertheless,
this modification is crucial. Firstly, it implies that
the spinor derivatives of $L$ with respect to $\theta, \bar\theta$, i.e.
$D_\alpha L$ and $\bar D_{\dot\alpha}L$, are shifted by
Grassmann parameters $\eta_\alpha, \bar\eta_{\dot\alpha}$ under $S$ supersymmetry
and so are Goldstone fermions for the considered linear realization of the
$N=2 \rightarrow N=1$ PBGS (the first component of $L$ is the Goldstone field for
the broken 5th translation with the generator $D$). Secondly, due to this modification,
the basic constraints of $N=2$ tensor multiplet \p{tmdef2} cease to be covariant and
should also be properly modified.

In order to find a proper deformation of \p{tmdef2}, one should construct
the covariant derivatives of the Goldstone superfield ${\cal L}^\omega$ by the standard
methods of nonlinear realizations \cite{nonl}. First, one calculates the Cartan
forms
\be
g^{-1}dg=i\omega_P^{\alpha\dot\alpha}P_{\alpha\dot\alpha}+i\mu\cdot Q+
 i{\bar\mu}\cdot{\bar Q} +i\nu\cdot S+i{\bar\nu}\cdot{\bar S}+i\omega_D D+
i\omega_M\cdot M+
 i{\bar\omega}_M\cdot {\bar M}\;,
\ee
\bea
&& \omega_P^{\alpha\dot\alpha}= -dx^{\alpha\dot\alpha}+
 id{\bar\theta}^{\dot\alpha}\theta^{\alpha}+id\theta^\alpha{\bar\theta}^{\dot\alpha}+
 id{\bar\xi}^{\dot\alpha}\xi^{\alpha}+id\xi^\alpha{\bar\xi}^{\dot\alpha}\;,\;
\nu^\alpha=d\xi^{\alpha}\;, \; {\bar\nu}^{\dot\alpha}=d{\bar\xi}^{\dot\alpha} \;,\nn
&& \mu^\alpha =  d\theta^\alpha+{\bar \xi}_{\dot\alpha}
 dx^{\alpha\dot\alpha}\; , \;
{\bar\mu}^{\dot\alpha} =  d{\bar\theta}^{\dot\alpha}+
  { \xi}_{\alpha}
 dx^{\alpha\dot\alpha}\; , \;
\omega_M^{\alpha\beta}=\bar\omega_M^{\dot\alpha\dot\beta}=0 \;,  \nn
&& \omega_D= d{\cal L}^\omega +i\left( \xi\cdot d\theta -{\bar\xi}\cdot d{\bar\theta}\right).
\eea
Next one defines the covariant derivatives of ${\cal L}^\omega$ as
\be
d{\cal L}^\omega = -\omega_P^{\alpha\dot\alpha} \partial_{\alpha\dot\alpha}{\cal L}^\omega
+\mu^\alpha {\Dt}{}^\theta_\alpha {\cal L}^\omega-{\bar\mu}^{\dot\alpha}
\bar{\Dt}^\theta_{\dot\alpha} {\cal L}^\omega+
\nu^\alpha \Dt^\xi_\alpha {\cal L}^\omega-
{\bar\nu}^{\dot\alpha} {\bar\Dt}^\xi_{\dot\alpha} {\cal L}^\omega \;,
\ee
where
\be\label{covder1}
{\Dt}{}^\theta_\alpha {\cal L}^\omega= D_\alpha {\cal L}^\omega +i\xi_\alpha \;, \;
\bar{\Dt}^\theta_{\dot\alpha} {\cal L}^\omega= \Db_{\dot\alpha} {\cal L}^\omega -i\xib_{\dot\alpha} \;, \;
\Dt^\xi_\alpha {\cal L}^\omega= \nabla_\alpha {\cal L}^\omega \;, \;
{\bar\Dt}^\xi_{\dot\alpha} {\cal L}^\omega = {\bar\nabla}_{\dot\alpha} {\cal L}^\omega \;.
\ee
Thus, only the spinor derivatives with respect to $\theta, \bar\theta$ get modified. Once again,
this modification affects only the superfield $L$ in ${\cal L}^\omega$, while the covariant
derivatives of $F, \bar F$ retain their previous form.

Now one can write the covariant constraints defining the $N=2$ tensor multiplet with the modified
transformations properties \p{qtr}, \p{str} by substituting the covariant derivatives \p{covder1}
in \p{htm5} or \p{tmdef2} for the previous ones. This gives
\be\label{tmdef2a}
\left\{ \begin{array}{l}
D_\alpha  F =0 \;, \\
D_\alpha L +i\xi_\alpha -\nabla_\alpha  F =0 \;, \\
\nabla_\alpha L + D_\alpha {\bar F} =0 \;, \\
\nabla_\alpha {\bar F}=0 \;,
\end{array}  \right. \qquad
\left\{ \begin{array}{l}
{\bar\nabla}_{\dot\alpha}  F =0 \;, \\
{\bar\nabla}_{\dot\alpha} L +\Db_{\dot\alpha}  F =0 \;, \\
\Db_{\dot\alpha} L -i\xib_{\dot\alpha} - {\bar\nabla}_{\dot\alpha} {\bar F} =0 \;, \\
\Db_{\dot\alpha} {\bar F}=0~.
\end{array}  \right.
\ee
The $S$-supersymmetry transformations \p{ssusy} of the $N=2$ tensor multiplet
${\cal L}^\omega$ induce the following transformations for the $N=1$ superfield
components $L,F,\bar F$ (with the constraints \p{tmdef2a} taken into
account)
\bea\label{n1tr}
&& \delta L = -i\left( \eta^\alpha \theta_\alpha - {\bar\eta}_{\dot\alpha} \tb^{\dot\alpha} \right)+
  \eta^\alpha D_\alpha {\bar F} - {\bar\eta}^{\dot\alpha} \Db_{\dot\alpha} F \;, \nn
&& \delta F = -\eta^\alpha D_\alpha L \;, \quad
\delta {\bar F} = {\bar\eta}^{\dot\alpha} \Db_{\dot\alpha} L \; .
\eea
They of course coincide with those given in \cite{BG2}.
The superfields $L$ and $F, \bar F$ are subjected to the same $N=1$ constraints \p{flatcon11}, which
remain covariant with respect to the modified transformations \p{n1tr}.
The entire set of Goldstone fields (the goldstino for the $N=2 \rightarrow N=1$ breaking and
the Goldstone field for the broken $D$ translations) is now accommodated by the
Goldstone $N=1$ superfield $L$.

It is evident from \p{n1tr} that one can construct an invariant `action' as follows
%with respect $N=2$ supersymmetry:
\be\label{action1f}
S=\frac{1}{4}\int d^4x d^2 \tb F + \frac{1}{4}\int d^4x d^2 \theta {\bar F} \;.
\ee
In order to make it meaningful, one should express the chiral supermultiplet $F,\bar F$
in terms of the Goldstone tensor multiplet $L$ by imposing proper covariant
constraints. These additional constraints were simply guessed in \cite{BG2} and
later re-derived in \cite{RT} from the nilpotency conditions imposed
on the appropriate $N=2$ superfield. They read
\be\label{flatcon}
F=-\frac{D^\alpha L\; D_\alpha L}{2-D^2 {\bar F}} \; \quad
{\bar F}=-\frac{\Db_{\dot\alpha} L\; \Db^{\dot\alpha} L}{2-\Db^2 F}
\ee
and can be easily solved \cite{BG2,RT}
\be\label{sol1flat}
F=- \psi^2+\frac{1}{2}  D^2 \left[
 { {\psi^2{\bar\psi}^2}\over{1+\frac{1}{2}A+\sqrt{1+A+\frac{1}{4}B^2}}}
 \right],
\ee
where
\bea
&& \psi_\alpha \equiv D_{\alpha}L\;, \quad {\bar\psi}_{\dot\alpha}\equiv
 {\bar D}_{\dot\alpha}L \;, \nn
&& A=\frac{1}{2}\left( D^2{\bar\psi}^2+{\bar D}^2\psi^2\right),\;
 B=\frac{1}{2}\left( D^2{\bar\psi}^2-{\bar D}^2\psi^2\right).
\eea
Finally, the action \p{action1f} becomes
\be\label{action2f}
S= -\frac{1}{4}\int d^4xd^2\theta {\bar\psi}^2 -
\frac{1}{4}\int d^4xd^2{\bar\theta} \psi^2 +
\frac{1}{4}\int d^4xd^4\theta
{ {\psi^2}{\bar\psi}^2 \over{1+\frac{1}{2}A+\sqrt{1+A+\frac{1}{4}B^2}}} \;.
\ee
It is a nonlinear extension of the standard $N=1$ tensor multiplet action.
In the bosonic sector it gives rise to the static-gauge Nambu-Goto action for a
L3-brane in $d=5$ Minkowski space, with one physical scalar of $L$ being
the transverse brane coordinate and another one being represented
by the notoph field strength. After dualizing $L$ into a pair of conjugated
chiral and antichiral $N=1$ superfields
(the notoph strength is dualized into a scalar field), the PBGS form
of the static-gauge action of super 3-brane in $d=6$ is reproduced \cite{BG2}.

Now we would like to demonstrate that the constraints
\p{flatcon} which play a central role in deriving the action \p{action1f}
are intimately related to the 5-dimensional nature of the brane
under consideration. They can be derived from the requirement of 5-dimensional Lorentz covariance.

Indeed, in order to find a proper place for the basic superfield
${\cal L}^\omega$ in the coset space, the $N=2, d=4$ Poincar\'{e}
superalgebra has been extended by a central charge generator $D$,
eqs. \p{flatalgebra}. This generator can be treated as the generator
of translations in the 5th direction and the full automorphism algebra
of \p{flatalgebra} can be checked to be $so(1,4)$
(we ignore the $R$-symmetry $SU(2)$ automorphisms).
The $d=5$ Lorentz algebra $so(1,4)$ includes,
besides the $d=4$ Lorentz generators $M_{\alpha\beta},{\bar M}_{\dot\alpha\dot\beta}$,
an additional $d=4$ vector $K_{\alpha\dot\alpha}$ belonging to the coset $SO(1,4)/SO(1,3)$.
The full set of the additional commutation relations reads as follows:
\bea\label{addcom}
&&
 i\left[  M_{\alpha\beta},K_{\rho\dot\rho}\right] =
\ve_{\alpha\rho}K_{\beta\dot\rho}+\ve_{\beta\rho}K_{\alpha\dot\rho}\;,\;
i\left[  {\bar M}_{\dot\alpha\dot\beta},K_{\rho\dot\rho}\right] =
\ve_{\dot\alpha\dot\rho}K_{\rho\dot\beta}+\ve_{\dot\beta\dot\rho}K_{\rho\dot\alpha}\;,\nn
&&  i\left[  K_{\alpha\dot\alpha},K_{\beta\dot\beta}\right] =
 -\ve_{\alpha\beta}{\bar M}_{\dot\alpha\dot\beta}-
  \ve_{\dot\alpha\dot\beta}M_{\alpha\beta}\;, \;
  i\left[ D, K_{\alpha\dot\alpha}\right] = 2P_{\alpha\dot\alpha}\;,\;
 i\left[  P_{\alpha\dot\alpha},K_{\beta\dot\beta}\right] =
 \ve_{\alpha\beta}\ve_{\dot\alpha\dot\beta}D \;, \nn
&&
  i\left[ K_{\alpha\dot\alpha}, Q_{\beta}\right]=-\ve_{\alpha\beta}{\bar S}_{\dot\alpha}\;,\;
i\left[ K_{\alpha\dot\alpha},{\bar Q}_{\dot\beta}\right]=
    -\ve_{\dot\alpha\dot\beta}S_{\alpha}\;, \nn
&& i\left[ K_{\alpha\dot\alpha}, S_{\beta}\right]=\ve_{\alpha\beta}{\bar Q}_{\dot\alpha}\;,\;
i\left[ K_{\alpha\dot\alpha},{\bar S}_{\dot\beta}\right]=
    \ve_{\dot\alpha\dot\beta}Q_{\alpha}\;.
\eea

One can ask whether the linear realization of the $N=2 \rightarrow N=1$ PBGS
with the Goldstone superfield ${\cal L}^\omega$
constructed above is compatible with this extra $SO(1,4)$ covariance.
The $SO(1,4)/SO(1,3)$ transformations of the superspace coordinates
and superfield ${\cal L}^\omega$ can be
easily found from the left action of the group element
$g_0=  e^{ia^{\alpha\dot\alpha}K_{\alpha\dot\alpha}}$ on the coset element \p{g}
\bea\label{ktrf}
&&\delta x^{\alpha\dot\alpha}=2a^{\alpha\dot\alpha}{\cal L}^\omega+
  ia^{\beta\dot\alpha}\left( \theta_\beta\xi^\alpha+\theta^\alpha\xi_\beta  \right)+
 ia^{\alpha\dot\beta}\left( {\bar\theta}_{\dot\beta}\xib^{\dot\alpha}+
   {\bar\theta}^{\dot\alpha}\xib_{\dot\beta}  \right),\nn
&& \delta \theta^{\alpha}=-ia^{\alpha\dot\alpha} \xib_{\dot\alpha}\;,\;
\delta {\bar\theta}_{\dot\alpha}=-ia^{\alpha\dot\alpha}\xi_\alpha\;, \;
\delta \xi^{\alpha}=-ia^{\alpha\dot\alpha}{\bar\theta}_{\dot\alpha}\;,\;
\delta {\bar\xi}_{\dot\alpha}=-ia^{\alpha\dot\alpha}\theta_\alpha\;,\nn
&& \delta {\cal L}^\omega = a_{\alpha\dot\alpha} x^{\alpha\dot\alpha} \;.
\eea
Now one should verify whether the defining constraints \p{tmdef2a} are consistent
with the transformations \p{ktrf}.
An equivalent but simpler check consists in passing to the $N=1$ superfield components
$L, F$ and $\bar F$ and examining whether their $SO(1,4)$ transformations are
consistent with the constraints \p{flatcon11}.  The active form
of $SO(1,4)$ transformations of these $N=1$ superfields can be found from \p{ktrf}
with taking into account the $\omega$ dependence of ${\cal L}^\omega$ in \p{l}
\bea\label{ktrf1}
&& \delta^* L = a_{\alpha\dot\alpha}x^{\alpha\dot\alpha}- a^{\alpha\dot\alpha} \partial_{\alpha\dot\alpha}
 \left( L^2-2 F{\bar F} \right) +i a^{\alpha\dot\alpha}\theta_{\alpha}\Db_{\dot\alpha}F
  -i a^{\alpha\dot\alpha}\tb_{\dot\alpha} D_{\alpha}{\bar F} \; ,\nn
&& \delta^* F=-2 a^{\alpha\dot\alpha}\partial_{\alpha\dot\alpha}\left( FL\right) +
    ia^{\alpha\dot\alpha}\tb_{\dot\alpha}D_{\alpha} L\; , \nn
&& \delta^* {\bar F}=-2 a^{\alpha\dot\alpha}\partial_{\alpha\dot\alpha}\left( {\bar F}L\right) -
    ia^{\alpha\dot\alpha}\theta_{\alpha}\Db_{\dot\alpha} L\; .
\eea
Note that these transformations are already essentially nonlinear, compared with the $S$- and $D$
transformations. Then the chirality of $F,\bar F$ implies
\be\label{addc1}
D_\alpha \bar F = \bar D_{\dot\alpha} F =0 \quad \Rightarrow \quad
\partial_{\alpha\dot\alpha}\left(F D_\beta L \right) =
\partial_{\alpha\dot\alpha}\left({\bar F} \Db_{\dot\beta}  L \right) =0 \; ,
\ee
while the $N=1$ tensor multiplet constraint yields one more condition
\be\label{addc2}
D^2 L=0 \quad \Rightarrow \quad  a^{\alpha\dot\alpha}\partial_{\alpha\dot\alpha} \left[ D^2 \left( L^2 -2 F{\bar F}\right)
  +4 F \right]=0 \;.
\ee
These conditions obviously cannot be satisfied with independent
$N=1$ superfields $L$ and $F, \bar F$. At the same time, it is straightforward to see
that the constraints \p{flatcon} solve both \p{addc1} and \p{addc2}. Thus the implementation
of the $d=5$ $SO(1,4)$ Lorentz covariance can be achieved only provided we
impose the constraint \p{flatcon} on our $N=2$ tensor multiplet, i.e. within the nonlinear
realization framework. This observation is crucial for the AdS case, where
all generators of the automorphism group appear in the anticommutators of $Q$ and $S$
supersymmetries. This case is the subject of the next Section.

\setcounter{equation}0
\section{AdS$_5$ background: Goldstone N=1 improved tensor multiplet}
In the previous Section we have shown that the implementation of the automorphism $SO(1,4)$
symmetry in the framework of a `linear' realization of spontaneously broken $N=2$
supersymmetry puts additional strong constraints on the Goldstone $N=2$ tensor
multiplet, giving rise to the genuine nonlinear realization in terms of the Goldstone
$N=1$ tensor multiplet pioneered in \cite{BG2}. Here we exploit
this observation, in order to construct a nonlinear realization describing the partial $1/2$ breaking
of the simplest AdS$_5$ supersymmetry $SU(2,2|1)$, with a suitable generalization of the $N=1$
tensor multiplet as a Goldstone one.

The superalgebra $su(2,2|1)$ contains a $so(2,4)\times u(1)$
bosonic subalgebra with generators $\left\{ P_{\alpha\dot\alpha}, M_{\alpha\beta},{\bar M}_{\dot\alpha\dot\beta},
K_{\alpha\dot\alpha},D\right\}$ and $\left\{ J\right\}$ and eight supercharges
$\left\{ Q_\alpha,{\bar Q}_{\dot\alpha},S_\alpha,{\bar S}_{\dot\alpha}\right\}$. It can be
considered either as a $N=1$ superconformal algebra in $d=4$ or as the superisometry of superspaces
with the AdS$_5$ or AdS$_5\times S^1$ bosonic bodies (depending on whether the $\gamma_5$ generator
$J$ is placed in the stability subgroup or in the coset).
We choose the basis in a such way, that the generators $K_{\alpha\dot\alpha}$ form a $so(1,4)$
subalgebra together with the $d=4$ Lorentz generators
$\left\{ M_{\alpha\beta},{\bar M}_{\dot\alpha\dot\beta}\right\}$, as in the first two
lines of \p{addcom}. The remaining non-trivial commutators read:
\bea
\label{adsalgebra}
&& i\left[  M_{\alpha\beta},P_{\rho\dot\rho}\right] =
\ve_{\alpha\rho}P_{\beta\dot\rho}+\ve_{\beta\rho}P_{\alpha\dot\rho}\;,\;
 i\left[  {\bar M}_{\dot\alpha\dot\beta},P_{\rho\dot\rho}\right] =
\ve_{\dot\alpha\dot\rho}P_{\rho\dot\beta}+\ve_{\dot\beta\dot\rho}P_{\rho\dot\alpha}\;,\nn
&& i\left[ D, P_{\alpha\dot\alpha}\right] = m P_{\alpha\dot\alpha}\;, \;
 i\left[ D, K_{\alpha\dot\alpha}\right] = 2P_{\alpha\dot\alpha}-m K_{\alpha\dot\alpha}\;,\nn
&&  i\left[  P_{\alpha\dot\alpha},K_{\beta\dot\beta}\right] =
 \ve_{\alpha\beta}\ve_{\dot\alpha\dot\beta}D -\frac{m}{2}\left(
 \ve_{\alpha\beta}{\bar M}_{\dot\alpha\dot\beta}+
  \ve_{\dot\alpha\dot\beta}M_{\alpha\beta}\right). \nn
&& \left\{ Q_{\alpha},S_{\beta} \right\}=-\ve_{\alpha\beta}\left( D+imJ \right)+
m M_{\alpha\beta}\;,\;
 \left\{ {\bar Q}_{\dot\alpha},{\bar S}_{\dot\beta} \right\}=-\ve_{\dot\alpha\dot\beta}
\left( D-imJ\right)+m {\bar M}_{\dot\alpha\dot\beta}\;,\nn
&& \left\{Q_\alpha, \bar Q_{\dot\alpha}\right\} = 2P_{\alpha\dot\alpha}~, \quad
\left\{S_\alpha, \bar S_{\dot\alpha}\right\} = 2P_{\alpha\dot\alpha} - 2m K_{\alpha\dot\alpha}~, \nn
&& i\left[ M_{\alpha\beta}, Q_{\gamma}\right]=\ve_{\alpha\gamma}Q_{\beta}+
  \ve_{\beta\gamma}Q_{\alpha} \equiv \left( Q\right)_{\alpha\beta,\gamma}\;,\;
 i\left[ M_{\alpha\beta}, S_{\gamma}\right]= \left( S\right)_{\alpha\beta,\gamma}\;,\nn
&& i \left[ D, Q_{\alpha}\right]=\frac{m}{2}Q_{\alpha}\;,\;
 i \left[ D, {\bar Q}_{\dot\alpha}\right]=\frac{m}{2}{\bar Q}_{\dot\alpha}\;,\;
 i \left[ D, S_{\alpha}\right]=-\frac{m}{2}S_{\alpha}\;,\;
 i \left[ D, {\bar S}_{\dot\alpha}\right]=-\frac{m}{2}{\bar S}_{\dot\alpha}\;,\nn
&&   \left[ J, Q_{\alpha}\right]=\frac{3}{2}Q_{\alpha}\;,\;
  \left[ J, {\bar Q}_{\dot\alpha}\right]=-\frac{3}{2}{\bar Q}_{\dot\alpha}\;,\;
  \left[ J, S_{\alpha}\right]=-\frac{3}{2}S_{\alpha}\;,\;
  \left[ J, {\bar S}_{\dot\alpha}\right]=\frac{3}{2}{\bar S}_{\dot\alpha}\;,\nn
&& i\left[ K_{\alpha\dot\alpha}, Q_{\beta}\right]=-\ve_{\alpha\beta}{\bar S}_{\dot\alpha}\;,\;
i\left[ K_{\alpha\dot\alpha},{\bar Q}_{\dot\beta}\right]=
    -\ve_{\dot\alpha\dot\beta}S_{\alpha}\;, \;
i\left[ K_{\alpha\dot\alpha}, S_{\beta}\right]=\ve_{\alpha\beta}{\bar Q}_{\dot\alpha}\;,\nn
&& i\left[ K_{\alpha\dot\alpha},{\bar S}_{\dot\beta}\right]=
    \ve_{\dot\alpha\dot\beta}Q_{\alpha}\;, \;
 i\left[ P_{\alpha\dot\alpha}, S_{\beta}\right]=m\ve_{\alpha\beta}{\bar Q}_{\dot\alpha}\;,\;
i\left[ P_{\alpha\dot\alpha},{\bar S}_{\dot\beta}\right]=
    m\ve_{\dot\alpha\dot\beta}Q_{\alpha}\;.
\eea
This provides an example of the `AdS basis' of conformal superalgebras
\cite{solvable,solvable1,dik,bik12,ikn} which
perfectly suits their interpretation as the superisometry groups of the appropriate
AdS superspaces. Indeed, the generators $P_{\alpha\dot\alpha}, D, J$ form
a maximal solvable bosonic subgroup in $su(2,2|1)$ and span the coset
$SO(2,4)/SO(1,4)\times U(1)\sim$ AdS$_5\times S^1$. The parameter $m$ has the meaning
of the inverse AdS$_5$ radius, $m= R^{-1}$. In the limit $m=0$ ($R = \infty$) \p{adsalgebra}
goes into the $N=1, d=5$ Poincar\'e superalgebra considered in the previous Section, with $D$
becoming the generator of translations along the 5th dimension. The generators $J$ and
$K_{\alpha\dot\alpha}, M_{\alpha\beta}, \bar M_{\dot\alpha\dot\beta}$ decouple
and generate outer $u(1)\oplus so(1,4)$ automorphisms.

Our goal is to construct an AdS$_5$ version of the Goldstone $N=2$ tensor supermultiplet
and then properly generalize the constraints \p{flatcon}. In terms of $N=1$
superfields this version is expected to involve some modification of the $N=1$ tensor multiplet $L$
and, as before, a pair of mutually conjugated $N=1$ chiral and anti-chiral
superfields $F, {\bar F}$. On these $N=1$ superfields we wish to realize an additional
supersymmetry, such that it forms, together with the manifest $N=1$ supersymmetry,
just the AdS$_5$  $SU(2,2|1)$ supersymmetry. Besides, in a close analogy
with the flat case, we wish that the following `action':
\be\label{action1}
S = \frac{1}{4}\int d^4x d^2{\bar\theta} F + \frac{1}{4}\int d^4x d^2 \theta {\bar F}
\ee
be an invariant of the AdS supersymmetry. Since the right-chiral integration measure
$d^4xd^2\bar\theta$ has a $D$ weight $-3m$ and,
with our normalization of $J$, a $U(1)$ charge $-3$, the superfield $F$
should carry $D$ and $J$ weights equal to $3m$ and $3$ ($\bar F$ has the same $D$ weight and
a $J$ charge equal to $-3$). Such an assignment will be a useful hint in finding out
the $SU(2,2|1)$ transformation laws of $L, F, \bar F$ and the appropriate constraints.

Our further strategy is similar, in its basic points, to what we did in the flat case.
The latter should be recovered as the $m=0$ limit of the $SU(2,2|1)$ construction,
what provides us with one more hint. We first define a non-standard coset realization of $SU(2,2|1)$
in a $N=2, d=4$ superspace by choosing $SO(1,3)\times U(1)$ as the stability subgroup
and associating two sets of Grassmann coordinates $(\theta^\alpha, \bar\theta^{\dot\alpha})$
and $(\xi^\alpha, \bar\xi^{\dot\alpha})$ with the generators $(Q_\alpha, \bar Q_{\dot\alpha})$
and $(S_\alpha, \bar S_{\dot\alpha})$. The coset parameters corresponding to the
$SO(2,4)/SO(1,4)$ generators $P_{\alpha\dot\alpha}$ and $D$ are the 4-coordinate
$x^{\alpha\dot\alpha}$ and the Goldstone $N=2$ superfield ${\cal L}^\omega(x,\omega,\theta,\xi)$.
We need not specify how the latter depends on $\omega$.
The only assertion to be exploited is that an analog of the inhomogeneously transforming
$N=2$ superfield $L$ of the flat case is still given by the $\omega$-independent
part of ${\cal L}^\omega$ (cf. \p{l})
\be\label{l2}
{\cal L}^\omega = L(x, \theta, \xi) + {\cal F}(\omega, {1\over \omega})~.
\ee
Thus, the supercoset element we start with is basically of the same form \p{g} as in the flat case.

The coset space techniques allow us to find the transformation properties
of the $N$ $=1$ superspace coordinates $( x^{\alpha\dot\alpha},\theta^\alpha,
\tb^{\dot\alpha})$ and the remaining coset parameters
$(\xi^\alpha, \bar\xi^{\dot\alpha}, {\cal L}^\omega)$ under $S$-supersymmetry. For our purposes,
it is enough to know them to zeroth order in $( \xi^\alpha,{\bar\xi}^{\dot\alpha} )$
\bea
&& \delta x^{\alpha\dot\alpha}=im\left( \eta_\beta x^{\beta\dot\alpha}\theta^\alpha+
  {\bar\eta}_{\dot\beta}x^{\alpha\dot\beta}\tb^{\dot\alpha}\right) +\frac{m}{2} \left(
  \tb{}^2\theta^\alpha {\bar\eta}^{\dot\alpha}+\theta^2 \eta^\alpha\tb^{\dot\alpha} \right), \nn
&& \delta \theta^\alpha= -m{\bar\eta}_{\dot\alpha}x^{\alpha\dot\alpha}
-im\left( \theta^2\eta^\alpha -
 \tb_{\dot\alpha}{\bar\eta}^{\dot\alpha} \theta^\alpha\right), \nn
&& \delta\tb^{\dot\alpha}=-m\eta_\alpha x^{\alpha\dot\alpha} +im\left(
  \tb^2 {\bar\eta}^{\dot\alpha} -\theta^\alpha \eta_\alpha \tb^{\dot\alpha} \right), \nn
&& \delta \xi^\alpha = \eta^\alpha \;, \quad \delta{\bar\xi}^{\dot\alpha}={\bar\eta}^{\dot\alpha}\;,
\label{adstr1} \\
&& \delta L= -i \left( \theta^\alpha \eta_\alpha -\tb_{\dot\alpha}{\bar\eta}^{\dot\alpha} \right).
\label{adsltr}
\eea
Thus, the active form of the $S$ transformation of the superfield $L$ (to zeroth order in $(\xi, {\bar \xi})$)
reads:
\be\label{adstrl1}
\delta^* L = -i \left( \theta^\alpha \eta_\alpha -\tb_{\dot\alpha}{\bar\eta}^{\dot\alpha} \right)
- \Delta x^{\alpha\dot\alpha} \partial_{\alpha\dot\alpha} L+
  \Delta\theta^\alpha D_{\alpha}L -\Delta{\bar\theta}^{\dot\alpha}
   {\bar D}_{\dot\alpha}L -\eta^\alpha \frac{\partial}{\partial \xi^\alpha}L
   - {\bar\eta}^{\dot\alpha}
\frac{\partial}{\partial {\bar\xi}^{\dot\alpha}}L \;,
\ee
where
\bea
&& \Delta x^{\alpha\dot\alpha}=2im\left( \eta_{\beta}x^{\beta\dot\alpha}
 \theta^{\alpha}+
 {\bar\eta}_{\dot\beta}x^{\alpha\dot\beta}{\bar\theta}^{\dot\alpha}\right)-
 m\left( \theta^2 \eta^\alpha{\bar\theta}^{\dot\alpha} -
 {\bar\theta}^2{\bar\eta}^{\dot\alpha}\theta^{\alpha}\right), \nn
&& \Delta \theta^\alpha=m{\bar\eta}_{\dot\alpha}x^{\alpha\dot\alpha}+
 im\left( \theta^2\eta^\alpha -
 {\bar\theta}_{\dot\alpha}{\bar\eta}^{\dot\alpha}\theta^{\alpha}\right)\;,\;
\Delta {\bar\theta}^{\dot\alpha} =m{\eta}_{\alpha}x^{\alpha\dot\alpha}-
 im\left( {\bar\theta}^2{\bar\eta}^{\dot\alpha} -
 {\theta}^{\alpha}{\eta}_{\alpha}{\bar\theta}^{\dot\alpha}\right),
\eea
and we should put $\xi^\alpha = \bar\xi^{\dot\alpha} = 0$ in both sides of \p{adstrl1}.

{}From the transformation law \p{adstrl1} one can guess the constraint
which replaces \p{flatcon11} in the AdS case
\be\label{adsconl}
\frac{1}{m}D^2 e^{-2mL}=\frac{1}{m}{\bar D}^2 e^{-2mL}=0\;.
\ee
Indeed, in the limit $m\rightarrow 0$, \p{adsconl} reduces to \p{flatcon11}.
If we neglect the last two terms in \p{adstrl1}, for the $N=1$ superfield $e^{-2mL}$
we just recover the standard superconformal transformation law of the
{\it improved} $N=1$ tensor multiplet \cite{impr}, with \p{adsconl} being covariant
under this realization. However, since we wish to gain a generalization of the
$N=2$ tensor multiplet transformation law \p{n1tr}, we cannot suppress such terms in
\p{adstrl1}. Similar terms should also modify the standard superconformal transformation
laws of the chiral $N=1$ superfields $F, \bar F$. In principle, we could fix
the $\xi, \bar\xi$ dependence of the $N=2$ superfields $L(x, \theta, \xi),
F(x,\theta, \xi), \bar F(x, \theta, \xi)$ entering ${\cal L}^\omega$,
by finding an analog of the covariant constraints \p{tmdef2a} for the present case.
However, in view of the highly non-trivial structure of the superalgebra \p{adsalgebra},
such a method would entail complicated technicalities. In fact, since we need
a realization of $SU(2,2|1)$ only on the $N=1$ superfield components
$L(x, \theta), F(x, \theta), \bar F(x, \theta)$,
it is simpler to guess the precise form of $(\partial L/\partial \xi^\alpha)|, \;
(\partial L/\partial \bar\xi^{\dot\alpha})|$ and similar contributions
to the transformation laws of $F, \bar F$ (here $\vert $ denotes retaining
$\xi, \bar\xi$ independent parts only). It turns out that the reasoning based on the $D$ and
$J$ weights, together with the requirement of compatibility
with the modified constraints \p{adsconl} and the chiral ones for $F, \bar F$,
fix these terms, up to a numerical coefficient. The latter is then determined by
the requirement that, in the flat $m=0$ limit, the transformations \p{n1tr}
are reproduced. One should take into account that $L$ is shifted by a
constant parameter under the action of the generator $D$, so $e^{-mL}$ has
a $D$ weight $m$ (spinor derivatives have a weight $m/2$).

In this way, we come to the following modified realization of conformal $S$ supersymmetry
on the $N=1$ superfields $L, F, {\bar F}$:
%(which are $\xi,{\bar \xi}$ independent components of
%$N=2$ superfields)
\bea\label{maintr}
\delta^* {\bar F} &=& 6im \theta^\alpha\eta_\alpha {\bar F} -
 \Delta x^{\alpha\dot\alpha} \partial_{\alpha\dot\alpha} {\bar F} +
 \Delta\theta^\alpha D_{\alpha} {\bar F} +
  ie^{-2mL}{\bar\eta}^{\dot\alpha}{\bar D}_{\dot\alpha} L \;, \nn
\delta^* F & =& -6im {\bar\theta}_{\dot\alpha}{\bar\eta}^{\dot\alpha} F -
  \Delta x^{\alpha\dot\alpha} \partial_{\alpha\dot\alpha} F -
 \Delta{\bar\theta}^{\dot\alpha} {\bar D}_{\dot\alpha} F +
  ie^{-2mL}{\eta}^\alpha D_\alpha L \;, \nn
\delta^* L& = & -i (\theta^\alpha\eta_\alpha -
 {\bar\theta}_{\dot\alpha}{\bar\eta}^{\dot\alpha}) -
 \Delta x^{\alpha\dot\alpha} \partial_{\alpha\dot\alpha} L+
  \Delta\theta^\alpha D_{\alpha}L -\Delta{\bar\theta}^{\dot\alpha}
   {\bar D}_{\dot\alpha}L \nn
&&  -ie^{2mL}\left[ \eta^\alpha D_\alpha\left( e^{2mL}\bF \right) +
  {\bar\eta}^{\dot\alpha}{\bar D}_{\dot\alpha}
       \left( e^{2mL}F \right)\right].
\eea
In the limit $m=0$, the transformations \p{maintr} go into \p{n1tr}. The compatibility
of the $F, \bar F$ transformation laws with the chirality conditions is just a consequence of
the constraint \p{adsconl}. The appearance of the weight pieces in
the transformations can be traced to the transformation properties
of the chiral integration measures in \p{action1}, and is required for
the invariance of \p{action1}. It is easy to find, e.g.
\be
\delta (d^4x_R d^2\bar\theta) =
\left(\frac{\partial \delta x_R^{\alpha\dot\alpha}}{\partial x_R^{\alpha\dot\alpha}} -
\frac{\partial \delta\bar\theta^{\dot\alpha}}{\partial \bar\theta^{\dot\alpha}}\right)
d^4x_rd^2\bar\theta =
6im\, (\bar\theta\cdot \bar\eta)\,d^4x_R d^2\bar\theta~,
\ee
where
\be
x^{\alpha\dot\alpha}_R = x^{\alpha\dot\alpha} -i \theta^\alpha\bar\theta^{\dot\alpha}~, \;\;
\delta x^{\alpha\dot\alpha}_R =
2im\, \bar\eta_{\dot\beta}x_R^{\alpha\dot\beta}\bar\theta^{\dot\alpha}~, \;
\delta \bar\theta^{\dot\alpha} = -m\, \eta_\alpha x_R^{\alpha\dot\alpha} +
im \, \bar\eta^{\dot\alpha} \bar\theta^2~.
\ee
Taking the transformations of $F, \bar F$ in a passive form, when only the first and
last terms in \p{maintr} are retained, we see that the weight pieces are necessary
for the invariance of \p{action1}.

The standard part of the transformation of $L$ in \p{maintr} is compatible with
the constraints \p{adsconl}, while in the additional piece (the last line of \p{maintr})
only the first and second terms manifestly obey the first and second constraints
in \p{adsconl}, respectively. Their sum, with $L, F, \bar F$ being independent $N=1$
superfields, is clearly inconsistent with \p{adsconl}. However, recall that the $SO(1,4)$
covariance in the flat case can be implemented, only provided the constraints \p{flatcon}
are imposed. In our case, the $SO(1,4)$ transformations appear in the closure of $Q$ and $S$
supersymmetries, so it is natural to expect that it is necessary to take into
account similar constraints already at the level of the $S$ transformations
\p{maintr} for self-consistency of the latter. It turns out that this is indeed the case.
We checked that the variations \p{maintr}
properly reproduce, in their closure with both themselves and manifest
$N=1$ supersymmetry, the remaining transformations of the AdS$_5$ superalgebra \p{adsalgebra},
only if $F,{\bar F}$ are subjected to the following nonlinear constraints:
\be\label{basiccon}
F=-{ {e^{-2mL}{D}^{\alpha}L{D}_{\alpha}L }\over
{2-e^{4mL} {D}^2 \bF}} \;, \;
{\bar F}=-{ {e^{-2mL}{\bar D}_{\dot\alpha}L{\bar D}^{\dot\alpha}L }\over
{2-e^{4mL} {\bar D}^2 F}} \;.
\ee
The latter constraints are compatible with both \p{adsconl} and the chirality properties of
$F, \bar F$. They go into \p{flatcon} at $m=0$. Once again, this modification
of \p{flatcon} can be easily guessed from the condition that $F$ and $\bar F$
possess $D$ weight $3m$ and $J$ charges $\pm 3$. Now, it is easy to check
that the full transformation of $L$ in \p{maintr} is compatible with the
constraint \p{adsconl} on the surface of \p{basiccon}. It is also a matter
of a straightforward computation, to check that
\p{basiccon} by themselves are covariant under the transformations \p{maintr}.
Thus, we see that, in the AdS$_5$ case, there are no direct analogs of either the
linear PBGS realization of the flat case, or the Goldstone tensor $N=2$
superfield. An analog of the latter can be consistently defined only on the surface of
the nonlinear constraints \p{basiccon}. Hence we are led, at once, to deal with
a nonlinear realization of $SU(2,2|1)$ in terms of the improved tensor
$N=1$ superfield $L(x,\theta)$.

Similarly to their flat counterparts, the constraint \p{basiccon} can be easily solved
\be\label{sol1}
F=-e^{-2mL}{\psi}^2+\frac{1}{2}{ D}^2 \left[
 { {\psi^2{\bar\psi}^2}\over{1+\frac{1}{2}A+\sqrt{1+A+\frac{1}{4}B^2}}}
 \right],
\ee
where
\bea
&& \psi_\alpha \equiv D_{\alpha}L\;, \quad {\bar\psi}_{\dot\alpha}\equiv
 {\bar D}_{\dot\alpha}L \;, \nn
&& A=\frac{1}{2}e^{2mL}\left( D^2{\bar\psi}^2+{\bar D}^2\psi^2\right)\;,\;
 B=\frac{1}{2}e^{2mL}\left( D^2{\bar\psi}^2-{\bar D}^2\psi^2\right).
\eea
Finally, the action \p{action1} can be written in the form
\be\label{action2}
S= -\frac{1}{4}\int d^4xd^2\theta e^{-2mL}{\bar\psi}^2 -
\frac{1}{4}\int d^4xd^2{\bar\theta} e^{-2mL}\psi^2 +
\frac{1}{4}\int d^4xd^4\theta
{ {\psi^2}{\bar\psi}^2 \over{1+\frac{1}{2}A+\sqrt{1+A+\frac{1}{4}B^2}}} \;,
\ee
or, equivalently, as
\be\label{action22}
S= \frac{1}{4}\int d^4xd^4\theta\left[{1\over m}L e^{-2mL} +
{ {\psi^2}{\bar\psi}^2 \over{1+\frac{1}{2}A+\sqrt{1+A+\frac{1}{4}B^2}}}\right].
\ee
The first term in \p{action22} is recognized as
the action of the improved tensor $N=1$ superfield \cite{impr}.

To see which sort of supersymmetric extended object the action \p{action22} describes let
us examine its bosonic core. Defining the bosonic components as follows:
\be
\phi=L|_{\theta=0}\;,\quad \left[D_{\alpha},{\bar D}_{\dot\alpha}\right]
e^{-2mL}|_{\theta=0}=-2mV_{\alpha\dot\alpha} \;,
\ee
where, in virtue of \p{basiccon},
\be
\partial_{\alpha\dot\alpha}V^{\alpha\dot\alpha}=0\;, \label{notoph}
\ee
the bosonic part of \p{action2} proves to be
\be
S_B=\int d^4xe^{-4m\phi}\left[ 1-\sqrt{1+\frac{1}{2}e^{6m\phi}V^2-2e^{2m\phi}
 (\partial \phi)^2 -
 e^{8m\phi} (V^{\alpha\dot\alpha}\partial_{\alpha\dot\alpha}\phi)^2} \label{bosL3}
 \right].
\ee
The latter is a conformally-invariant extension of the static gauge Nambu-Goto action for a
L3-brane in $d=5$: the dilaton $\phi$ can be interpreted as a radial brane coordinate, while
$V^{\alpha\dot\alpha}$ is the field strength of the notoph, and it contributes one more scalar
degree of freedom on shell. In the limit $V^{\alpha\dot\alpha}=0\,$, the action \p{bosL3} becomes just the
static-gauge form of the Nambu-Goto action for 3-brane on AdS$_5$ in the `solvable-subgroup'
parametrization \cite{solvable,dik}.

Thus we conclude that the superfield action \p{action22} describes the static-gauge AdS$_5$
super L$3$-brane which can be defined as a superconformally invariant generalization of the
flat superspace L$3$-brane of refs. \cite{BG2}-\cite{R2}. To our knowledge, such an action
was never given before. On the other hand, the flat superspace L$3$-brane action beyond the
static gauge was deduced in ref. \cite{howe} in the framework of superembedding approach. It would
be tempting to recover the action \p{action22} as a gauge-fixed form of some appropriate
worldvolume action in this approach.

We wish to point out that the Goldstone superfield action \p{action22} corresponds to the one-half
partial breaking of $SU(2,2|1)$ down to its $N=1, d=4$ super Poincar\'e subgroup which, together with
the $R$-symmetry (or $\gamma_5$) subgroup generated by $J$, are the only linearly realized symmetries
of this PBGS pattern. Likewise, the only linearly realized symmetry of the bosonic action \p{bosL3}
is the 4-dimensional worldvolume Poincar\'e symmetry.

The $U(1)$ $R$-symmetry generated by $J$ gets nonlinearly realized after performing
a duality transformation. As well known, the notoph field strength $V^{\alpha\dot\alpha}$ can be dualized into
an off-shell scalar, by first introducing the constraint \p{notoph} into the action
with a Lagrange scalar multiplier and then eliminating $V^{\alpha\dot\alpha}$,
using its algebraic equation of motion, in terms of the additional scalar field.
Extending \p{bosL3} as follows:
\be
S_B \quad \Rightarrow \quad S^{dual}_B = S_B + \int d^4x
\lambda \partial_{\alpha\dot\alpha}V^{\alpha\dot\alpha}
\ee
and eliminating $V^{\alpha\dot\alpha}$, after some algebra we obtain
\be\label{dual1}
S^{dual}_B = \int d^4x\,e^{-4m\phi}\left\{1 - \sqrt{1 -2e^{2m\phi}
\left[(\partial \phi)^2 + (\partial \lambda)^2\right] +
4e^{4m\phi}\left[(\partial\phi)^2(\partial\lambda)^2 -
(\partial\phi\partial\lambda)^2\right]}\right\}.
\ee
After passing to Cartesian $\mathbb{R}^2$ coordinates
\be
Z^1 = r\, \cos \vartheta~, \;\; Z^2 = r\, \sin \vartheta~, \;\; r \equiv e^{-m\phi}~, \;\;\vartheta
\equiv m\,\lambda~, \label{cartangular}
\ee
one can rewrite \p{dual1} in a nice form
\be\label{dual2}
S^{dual}_B = \int d^4 x \, |Z|^4\left[1 - \sqrt{-\mbox{det}\left(\eta_{\mu\nu} -
{2\over m^2}\frac{\partial_\mu Z^n\partial_\nu Z^n}{|Z|^4}\right)}\right],
\ee
which is just the $S^5 \rightarrow S^1$ reduction of the scalar part of the D3-brane
action on AdS$_5\times S^5$ \cite{ads,mt2}, i.e. the static-gauge
Nambu-Goto action of a scalar 3-brane on AdS$_5\times S^1$ (with AdS$_5$ and $S^1$ having
equal radii $\sim m^{-1}$). In the next Section we shall perform this duality transformation
at the full superfield level and obtain
a $SU(2,2|1)$ invariant action of the Goldstone chiral $N=1$ superfield which yields in
its bosonic sector just \p{dual1} or \p{dual2}. It will be argued there that
$\lambda(x)$ is the coset parameter associated with the $U(1)$ generator $J$, which
supports its interpretation as an angular variable
in \p{cartangular}.\footnote{Some appropriate periodicity conditions should be imposed
on $\lambda(x)$ to make such an interpretation correct.}
%the one anticipated in \cite{kum}.

\setcounter{equation}0
\section{Dual Goldstone superfield action on AdS$_5$ x S$^1$}
In order to carry out a duality transformation at the full superfield level, we begin with
the superfield action \p{action22} and relax the basic constraints \p{adsconl} by adding
a Lagrange multiplier
\be\label{relaxaction}
S^{dual}=\frac{1}{4}\int d^4 x d^2 \theta d^2 \tb \left[ -\frac{1}{2m^2} Y\left( \ln Y-1\right) +
  \frac{Y^{-4}}{(2m)^4} (D Y)^2 (\Db Y)^2 f+ \frac{Y}{2m}\left( \vfi+\bvfi\right) \right],
\ee
where
\be\label{deffi}
Y\equiv e^{-2mL}\;,\quad \Db_{\dot\alpha} \vfi= D_\alpha \bvfi=0 \; .
\ee
Next, we vary the action \p{relaxaction} with respect to $Y$, in order to obtain an
algebraic equation that would allow us to trade $Y$ for $\vfi, \bvfi$
\bea\label{eq1}
\frac{1}{2m^2}\ln Y -\frac{1}{2m}\left( \vfi + \bvfi\right) & = &  a_1 (D Y)^2 + a_2 (\Db Y)^2 +
  c_{\alpha\dot\alpha} D^{\alpha}Y \Db^{\dot\alpha} Y  \nn
&& + a_3^{\alpha}D_{\alpha}Y (\Db Y)^2+
   a_4^{\dot\alpha}\Db_{\dot\alpha}Y(D Y)^2 +a_5 (D Y)^2 (\Db Y)^2 \;,
\eea
where all terms in the r.h.s come from the variation of the four-fermion term in \p{relaxaction}.
The coefficients $a_n$ and $c_{\alpha\dot\alpha}$ are functions of $Y$ and its derivatives.

Plugging \p{eq1} back into the action \p{relaxaction} gives us the following expression:
\be\label{action23}
S^{dual}=\frac{1}{4}\int d^4 x d^2 \theta d^2 \tb  e^{m(\vfi+\bvfi)}\left[ \frac{1}{2m^2} +\left(
  \frac{Y^{-4}}{(2m)^4}f -2m^2 a_1 a_2 -\frac{m^2}{4} c^2 \right)(D Y)^2 (\Db Y)^2 \right].
\ee
Thus, we need to solve the rather complicated equation \p{eq1}, which expresses
$Y$ in terms of $\vfi,\bvfi$ only up to the second order in the fermions. Moreover,
the functions $a_1$ and $a_2$ are proportional to $\Db^2 Y$ and $D^2 Y$, respectively.
It was explicitly shown in \cite{R2} that such terms can be reabsorbed into a redefinition
of a chiral Lagrangian multiplier. Therefore we can discard such terms in the action.

To summarize, the equation we have to solve reads:
\be\label{eq2}
-\frac{1}{2m^2}\ln Y +\frac{1}{2m}\left( \vfi + \bvfi\right)  +
  c_{\alpha\dot\alpha} D^{\alpha}Y \Db^{\dot\alpha} Y =0 \; ,
\ee
where
\be\label{eq3}
c_{\alpha\dot\alpha}= \frac{Y^{-4}}{8m^4}\left[H{\bar G}_{\alpha\dot\alpha} -
{\bar H} G_{\alpha\dot\alpha}\right].
\ee
Here,
\be\label{defAB}
H= 2f+ \left( f_A+f_B\right) \left(A+B\right) \; , \quad
{\bar H}= 2f+ \left( f_A-f_B\right) \left(A-B\right),
\ee
and
\be\label{defG}
G_{\alpha\dot\alpha} \equiv D_\alpha \Db_{\dot\alpha} Y \;, \quad
{\bar G}_{\alpha\dot\alpha} \equiv \Db_{\dot\alpha} D_\alpha  Y .
\ee

Acting by two spinor derivatives on  eq. \p{eq2} and omitting terms with
fermions we obtain the following nonlinear equations:
\bea\label{eq4}
&& -2i\partial_{\alpha\dot\alpha}\vfi= \frac{Y^{-1}}{m}\left[
  \left( 1-\frac{1}{4}\left( A+B\right) {\bar H}\right) {\bar G}_{\alpha\dot\alpha} +
    \frac{1}{4} \left( A-B\right) H G_{\alpha\dot\alpha} \right], \nn
&& -2i\partial_{\alpha\dot\alpha}\bvfi=\frac{Y^{-1}}{m}\left[
  \left( 1-\frac{1}{4}\left( A-B\right)  H \right) G_{\alpha\dot\alpha} +
    \frac{1}{4} \left( A + B\right) {\bar H} {\bar G}_{\alpha\dot\alpha}\right].
\eea
Squaring these equations and taking their cross-product we find the relations
between $A,B$ and $P$ defined as
\be
P \equiv \frac{Y^{-3}}{2m^2} G^{\alpha\dot\alpha}{\bar G}_{\alpha\dot\alpha}\;,
\ee
on the one hand, and the corresponding duals
\bea\label{defabp}
&a=-e^{-m\left( \vfi+\bvfi\right)} \left[ \left( \partial\vfi\right)^2 +
   \left( \partial\bvfi\right)^2\right], \;
b=-e^{-m\left( \vfi+\bvfi\right)} \left[ \left( \partial\vfi\right)^2 -
   \left( \partial\bvfi\right)^2\right], &\nn
&p=-e^{-m\left( \vfi+\bvfi\right)} \partial^{\alpha\dot\alpha}\vfi
         \partial_{\alpha\dot\alpha}\bvfi \;,&
\eea
on the other. The relations
\bea\label{eqq}
&& a+b=(A-B)\left[ \left( 1-\frac{1}{4}{\bar H}(A+B)\right)^2+ \frac{1}{2}HP\left( 1-\frac{1}{4}{\bar H}(A+B)\right)
+\frac{1}{16}H^2(A^2-B^2)\right], \nn
&&a-b=(A+B)\left[ \left( 1-\frac{1}{4}H(A-B)\right)^2+ \frac{1}{2}{\bar H}P\left( 1-\frac{1}{4}H(A-B)\right)
+\frac{1}{16}{\bar H}^2(A^2-B^2)\right], \nn
&& a+p=A+P
\eea
can be exactly solved, yielding the following expressions:
\be\label{soleqq}
A=\frac{2(a^2-b^2+a(2+p))}{b^2-a^2+(p+2)^2}\;, \;
B=-\frac{2b}{\sqrt{b^2-a^2+(p+2)^2}}\;,\; P=a+p-A\;.
\ee

As a last step, we should express $(DG)^2(\Db G)^2$ in terms of spinor derivatives of
$\vfi,\bvfi$. Once again, acting on the basic equation \p{eq1} with
the derivatives $D_{\alpha}$ and $\Db_{\dot\alpha}$
and keeping only terms linear in the fermions, we find
\be\label{eq4f}
D_{\alpha}\vfi=\frac{1}{m}{D_\alpha \ln Y}+ 2m c^{\beta\bar\beta}G_{\alpha\dot\beta}D_{\beta}Y \;,\;
\Db_{\dot\alpha}\bvfi=\frac{1}{m}{\Db_{\dot\alpha} \ln Y}-
2m c^{\beta\bar\beta}{\bar G}_{\beta\dot\alpha}\bar D_{\dot\beta}Y \;.
\ee
{}From \p{eq4f} one can deduce
\be\label{eq4fa}
(D G)^2(\Db G)^2= \frac{ m^4 Y^4 (D\vfi)^2(\Db \bvfi)^2}{
\left( 1 +2m^2 Y c_{\alpha\dot\alpha}G^{\alpha\dot\alpha}+m^4 Y^2 c^2 G^2\right)
\left( 1 -
 2 m^2 Y c_{\alpha\dot\alpha}{\bar G}^{\alpha\dot\alpha}+m^4 Y^2 c^2 {\bar G}^2\right)}\;.
\ee

Finally, plugging all this in the actions \p{action23}, we find the dual action in
a rather simple form
\bea\label{dualS}
S^{dual}&=&\frac{1}{8}\int d^4 x d^4\theta \left( \frac{1}{m^2}e^{m(\vfi+\bvfi)} \right. \\
&&+\left .\frac{\frac{1}{8} (D\vfi)^2(\Db \bvfi)^2}{1-e^{-m(\vfi+\bvfi)}\partial\vfi\partial\bvfi +
\sqrt{ (1-e^{-m(\vfi+\bvfi)}\partial\vfi\partial\bvfi)^2 -e^{-2m(\vfi+\bvfi)}
  (\partial\vfi)^2(\partial\bvfi)^2}} \right).\nonumber
\eea
This action goes into the flat $N=2 \rightarrow N=1$ chiral Goldstone superfield action
of \cite{RT,R2} in the limit $m=0$ and is obviously $SU(2,2|1)$ invariant, as it was
obtained by dualizing the $SU(2,2|1)$ invariant action \p{action2}, \p{action22}.

In principle, one can find the precise form of the $SU(2,2|1)$ transformations of the chiral
Goldstone superfields $\vfi$, $\bvfi$, but we do not give them here, because
they look not too illuminating. We only note that the standard $U(1)$ isometry associated with
the duality transformation, viz. $\delta\vfi = i\alpha$, $\delta\bvfi = -i\alpha$, now appears
in the closure of
$Q$ and $S$ transformations of the Goldstone superfields, with the imaginary part of $\vfi|$
as the corresponding Goldstone field. Hence, it is just the $J$ (or $\gamma_5$) symmetry of $SU(2,2|1)$.
In other words, performing a duality transformation brings this symmetry from the stability subgroup into the
coset. Actually, this can be seen already at the level of dualization of the standard improved
$N=1$ tensor supermultiplet, which corresponds to the approximation of neglecting the last line
in the transformation law of $L$ in \p{maintr} and keeping only the first term in the action \p{action22}.
In this case, it is easy to find the precise expression of $Y$ in terms of $\vfi, \bvfi$
\be
Y = e^{m(\vfi+\bvfi )}\;.
\ee
Then, the standard $S$ transformation law of $Y$ is reproduced by the following
transformations of chiral superfields
\be
\delta^* \vfi = 2i \theta^\alpha\eta_\alpha  -
 \Delta x^{\alpha\dot\alpha} \partial_{\alpha\dot\alpha} \vfi +
 \Delta\theta^\alpha D_{\alpha} \vfi \;, \;
\delta^* \bvfi  = -2i {\bar\theta}_{\dot\alpha}{\bar\eta}^{\dot\alpha} -
  \Delta x^{\alpha\dot\alpha} \partial_{\alpha\dot\alpha} \bvfi -
 \Delta{\bar\theta}^{\dot\alpha} {\bar D}_{\dot\alpha} \bvfi \;.\label{fitrans}
\ee
It is easy to see that, in the closure of this transformation with the standard $N=1$ Poincar\'e
supersymmetry, there appears a complex bracket parameter. Its real and imaginary parts
are just the dilatonic and $\gamma_5$ weight transformations. This property persists
in the complete nonlinear version of \p{fitrans}. A similar
`jumping' of the $R$-symmetry generator from the stability subgroup to the coset after a duality
transformation was observed in \cite{ikl}, in the study of the relation between real
and complex forms of $N=2$ superconformal mechanics associated with the nonlinear
realization of the $N=2, d=1$ superconformal group $SU(1,1|1)\sim OSp(2|2)$.

The bosonic part of the action \p{dualS} reads:
\be\label{dualSb}
S^{dual}_B=\int d^4 x e^{2m(\vfi+\bvfi)}\left[1+\sqrt{ \left( 1-e^{-m(\vfi+\bvfi)} \partial\vfi\partial\bvfi\right)^2
  - e^{-2m(\vfi+\bvfi)}(\partial\vfi)^2 (\partial\bvfi)^2}\right]
\ee
and it coincides with \p{dual1}, after the following identifications:
\be
\phi= -\frac{1}{2}\left( \vfi+\bvfi \right), \quad \lambda =
\frac{i}{2}\left( \vfi - \bvfi \right).
\ee

Thus, we conclude that the Goldstone superfield action \p{dualS} describes a situation where
$SU(2,2|1)$ is nonlinearly realized in its coset over the subgroup $SO(1,3)$, with a $N=1, d=4$
Poincar\'e supersymmetry realized in the standard linear way on $N=1$ superspace coordinates and
Goldstone superfields. The $S$ supersymmetry is broken, along with the $D$, $J$ and
$SO(1,4)/SO(1,3)$ $K_{\alpha\dot\alpha}$ transformations (the `Goldstone field'
for the latter is basically the $x$-derivative of the dilaton). The independent bosonic
Goldstone fields and $x^{\alpha\dot\alpha}$ parameterize the coset manifold
AdS$_5\times S^1 \propto \{x^{\alpha\dot\alpha},\phi\}\otimes \{\lambda\}$.
The bosonic part of the action \p{dualS} is just the static-gauge Nambu-Goto action of
3-brane on the latter manifold.

This solves the problem of constructing a minimal Goldstone superfield action
for the PBGS option considered.

\setcounter{equation}0
\section{Discussion}
In this paper we have constructed new nonlinear realizations of
the simplest AdS$_5$ superisometry group (viz. $N=1, d=4$ superconformal group) $SU(2,2|1)$,
in terms of a $N=1, d=4$ improved tensor and chiral Goldstone superfields. We have set up the minimal
Goldstone superfield action for the first option, by generalizing the approach applied
earlier to the case of flat Minkowski background. This generalization is not straightforward,
and it essentially relies on a novel interpretation of the basic constraints of refs. \cite{BG2,RT,R2},
as a guarantee of hidden $d=5$ $SO(1,4)$ covariance. The minimal Goldstone superfield action
for the second PBGS option has been obtained by dualizing the action for the first one.
The actions constructed contain no free parameters (as opposed to the candidate Goldstone
superfield action of ref. \cite{kum}). They provide a manifestly $N=1$ supersymmetric off-shell
superfield form of the worldvolume actions for a L3-superbrane on AdS$_5$ and a scalar super 3-brane
on AdS$_5\times S^1$, respectively. The latter is a truncation
of the action for a super AdS$_5\times S^5$ D3-brane.
In the limit of infinite AdS$_5$ radius, the new actions go into their flat superspace
counterparts which describe the partial breaking of $N=2, d=4$ supersymmetry down
to $N=1$ supersymmetry \cite{BG2,RT,R2,BG1}. Similarly to the flat superspace Goldstone
superfield Lagrangians, their AdS$_5$ analogs do not behave under
$SU(2,2|1)$ supersymmetry transformations as tensors. Rather, they are shifted
by a total derivative. In this respect, they are reminiscent of the WZW or CS Lagrangians.

The study in this paper, together with the results of \cite{dik}, can be regarded as
first steps in a program of constructing off-shell Goldstone superfield actions for
various patterns of partial breaking of AdS$\times S$ supersymmetries and their
non-trivial contractions corresponding to pp-wave type backgrounds. One of the obvious
related tasks (as already mentioned in \cite{kum}) is the quest for an action
corresponding to the half-breaking of the $N=2$ AdS$_5$ supergroup $SU(2,2|2)$, in a supercoset
with the AdS$_5\times S^1$ bosonic part. In this case, the basic Goldstone superfield
that we expect to deal with should be the appropriate generalization
of the $N=2$ Maxwell superfield strength. The relevant minimal action
should be a superconformally invariant version of the Dirac-Born-Infeld action describing
$N=4 \rightarrow N=2$ partial breaking in flat superspace \cite{bik11,kt}.
Note that in the flat case there exists one more
$N=2 \rightarrow N=1$ PBGS option associated with the choice of a vector $N=1, d=4$
multiplet as a Goldstone one and corresponding to a space-filling
$N=1$ D3-brane \cite{BG3}. No AdS$_5$ analog of this realization exists.
The reason is that the $SU(2,2|1)$ invariance requires the presence of a dilaton field in the
relevant $N=1$ Goldstone supermultiplet.
In the Goldstone $N=2$ vector supermultiplet there are two scalar fields and,
therefore, the above objection can be circumvented.

Another interesting problem is to extend the `holographic map' of ref. \cite{bik12}
to the superconformal  PBGS cases, including those studied in the present paper.
We expect the existence of a nonlinear change of Goldstone superfields and
$N=1$ superspace coordinates which maps the nonlinear realization \p{maintr}
(and its counterpart for the chiral Goldstone superfields) onto the standard
nonlinear realization of $SU(2,2|1)$ regarded as a $N=1, d=4$ superconformal group \cite{nonl1}.
The minimal Goldstone superfield actions constructed above are expected to be mapped onto
some non-linear higher-derivative extensions of the standard $N=1, d=4$ superconformal actions
of the improved tensor and chiral $N=1$ superfields used as the Goldstone ones for
the standard nonlinear realizations of $SU(2,2|1)$, similarly to what takes place for
bosonic AdS actions \cite{bik12}.

\section*{Acknowledgements}
This work was partially supported by INTAS grant No 00-0254 and the Iniziativa
Specifica MI12 of the INFN Commissione Nazionale IV. The research of S.B. is supported in part by
European Community's Human Potential Programme contract HPRN-CT-2000-00131 and
NATO Collaborative Linkage Grant PST.CLG.979389. The research of E.I. and S.K. is supported in part by grants DFG No.436
RUS 113/669, RFBR-DFG 02-02-04002, RFBR-CNRS 01-02-22005 and a grant of the Heisenberg-Landau
program. E.I. and S.K. thank INFN-LNF for warm hospitality during the
course of this work.

\end{document}